# Hyper-sampling imaging

Hemeng Xue[1,2,3]†, Mingtao Shang[1,2,3]†, Ze Zhang[1,2,3,6]*†, Hongfei Yu[1,2,3], Jinchao Liang[1], Meiling Guan[1,2,6], Chengming Sun[1], Huahua Wang[1,5], Shufeng Wang[2], Zhengyu Ye[1], Feng Gao[4], Lu Gao[5]

[1]Aerospace Information Research Institute, Chinese Academy of Sciences; Beijing, 100094, China.

[2]College of Photonics and Optical Engineering, Aerospace Information Technology University; Jinan, 250299, China.

[3]Qilu Aerospace Information Research Institute; Jinan, 250100, China.

[4]MOE Key Laboratory of Weak-Light Nonlinear Photonics, TEDA Applied Physics Institute and School of Physics, Nankai University; Tianjin, 300457, China.

[5]China University of Geosciences (Beijing); Beijing, 100083, China.

[6]Key Laboratory of Computational Optical Imaging Technology, Chinese Academy of Sciences; Beijing, 100094, China.

†These authors contributed equally to this work.

*Corresponding author. Email: zhangze@aircas.ac.cn

# **Abstract**

The transition from optical film to digital image sensors (DIS) in imaging systems has brought great convenience in human life. However, the sampling resolution of DIS is considerably lower than that of optical film due to the limitation that the pixels are significantly larger than the silver halide molecules. How to break DIS's sampling limit and achieve high-resolution imaging is highly desired for imaging applications. In our research, we have developed a novel mechanism that allows for a significant reduction in the smallest sampling unit of DIS to as small as 1/16th of a pixel, or even smaller, through measuring the intra-pixel quantum efficiency for the first time and recomputing the image — a technique we refer to as hyper-sampling imaging (HSI). Employing the HSI method, the physical sampling resolution of regular DIS can be enhanced by $4 \times 4$ times or potentially higher, and detailed object information can be acquired. The HSI method has undergone rigorous testing in real-world imaging scenarios, demonstrating its robustness and efficiency in overcoming the sampling constraints of conventional DIS. This advancement is particularly beneficial for applications such as remote sensing, long-range reconnaissance, and astronomical observations, where the ability to capture fine details is paramount.

# **Introduction**

Over the past decades, since digital image sensor (DIS) replaced the usage of film in numerous fields, it has brought groundbreaking advancements in various applications, such as remote sensing, mobile photography, among others[1-3]. However, the current technology level of micro-nano fabrication imposes significant constraints, making it nearly unfeasible to produce DIS with pixel dimensions, scale, and response uniformity that rival the precision of film. The smallest imaging unit in film, the silver halide molecule or cluster, sets a high standard that modern DIS struggle to match, thereby placing substantial limitations on the evolution of contemporary imaging technology[4-6]. This limitation has raised concerns and prompted questions about the comparative image quality. For example, critiques have highlighted that the imagery from the Chang'e-4 lunar mission, launched in 2018, appears to be of lower quality than the photographs taken by the Apollo missions nearly six decades prior[7-10].

To compensate the shortage of DIS, researchers have developed at least three kinds of approaches to enhance its performance. The first kind involves refining manufacturing techniques to decrease pixel size and increase the pixel count on image sensors[11-13]. Nonetheless, these methods are encountering diminishing returns due to factors such as pixel crosstalk, readout speed, and escalating production costs, which are pushing them towards technical ceilings and making them progressively challenging to implement in practical applications[14,15]. The second kind employs image processing algorithm based on deep learning or interpolation techniques[16-19]. While these solutions are adept at enhancing the visual appeal of images, they often fall short of transcending the physical constraints of DIS to capture high-frequency information intrinsic to the details of objects being imaged. The third kind involves the precise calibration of the image sensor to accurately determine its pixel response function[20-26]. However, due to the inherent photoelectric response mechanisms of pixels, these methods struggle to extract intra-pixel details, thereby failing to surpass the sensor's sampling limit and acquire additional object information in a tangible, physical sense[27].

In this paper, we present an innovative mechanism and methodology that transcends the inherent physical sampling limit of DIS through the process of measuring intra-pixel quantum efficiency (*QE*) and re-computing the image, a technique we refer to as hyper-sampling imaging (HSI)[28,29]. To the best of our knowledge, this is the first time the intra-pixel *QE* has been measured, which may bring many possibilities for development in the areas of image quality enhancement, image sensor manufacturing and etc. Our research, validated through laboratory and outdoor experiments, confirms the hyper-sampling effect of HSI, showing that each single pixel can be used as 16 sampling units, or potentially more. The HSI method has been proved to be highly robust and efficient to improve both the resolution and the quality of images using regular DIS in practical applications, and might be a breakthrough for modern imaging technology.

## **Results**

**Experimental measurement of intra-pixel quantum efficiency**

In experiment, we used a CCD (C14041-10U, 320 × 256 pixels, Hamamatsu, Japan) as the target DIS to measure its intra-pixel *QE* and realize HSI. We assumed that each pixel is divided into $k \times k$ subregions, in which $k = 1, 2, 3…$ is the number of subregions along transverse or longitudinal edge. A series of time-varying interference patterns of two laser beams $g_1(\omega)$ and $g_2(\omega + \Delta\omega)$ are used as the input optical field, where $\Delta\omega$ is their frequency difference, and is set to be lower than 10 Hz in order to be well sampled by the DIS. For a specific time $t_\eta$, the numerical expression $I_{\eta,m,n}$ of interference patterns' image can be read out from CCD, while the exact expression $S_{\eta,m,n}(i, j)$ of input intensity field can be fitted out in the time domain, in which $i = 1$, 2… $k$; $j = 1, 2… k$ is the serial number of subregions for a pixel ($m$, $n$). By subscribing the $S_{\eta,m,n}(i, j)$ and $I_{\eta,m,n}$ into equations, the intra-pixel $QE$ across the whole CCD plane can be solved out (Fig. 1A and supplemental materials).

The measured $QE$ of the whole CCD and a randomly selected pixel (200, 260) for $k = 1, 3, 4$ under different light intensities is shown in Fig. 1B and C. We can see that the $QE$ vary not only with the spatial location among pixels, but also among subregions inside each pixel. To observe the $QE$ variation trend among pixels across CCD plane under different light intensities, we purposively selected pixels (1, 1), (1, 318), (255, 1), (255, 318) located at the corner of the CCD, and pixels (126, 155), (126, 160), (128, 155), (128, 160) located at the center of the CCD, and drew their $QE$ curves according to $S_{\eta,m,n}(i, j)$ in the same coordinate, as shown in the left figure of Fig. 1C. To investigate the $QE$ variation trend inside pixel under different light intensities, we take pixel (82, 151) as an example and consciously drew the $QE$ values of its subregions (1, 4, 13, 16) located at the corner, and subregions (6, 7, 10, 11) located at the center under different $S_{\eta,m,n}(i, j)$, as shown in the right figure of Fig. 1C. From the curves, we can see that both the $QE$ values of pixels and subregions vary according to the $S_{\eta,m,n}(i, j)$. To physically improve the resolution and quality of images, it is essential to know this $QE$ variation across the CCD plane since this $QE$ contains the display error of DIS.

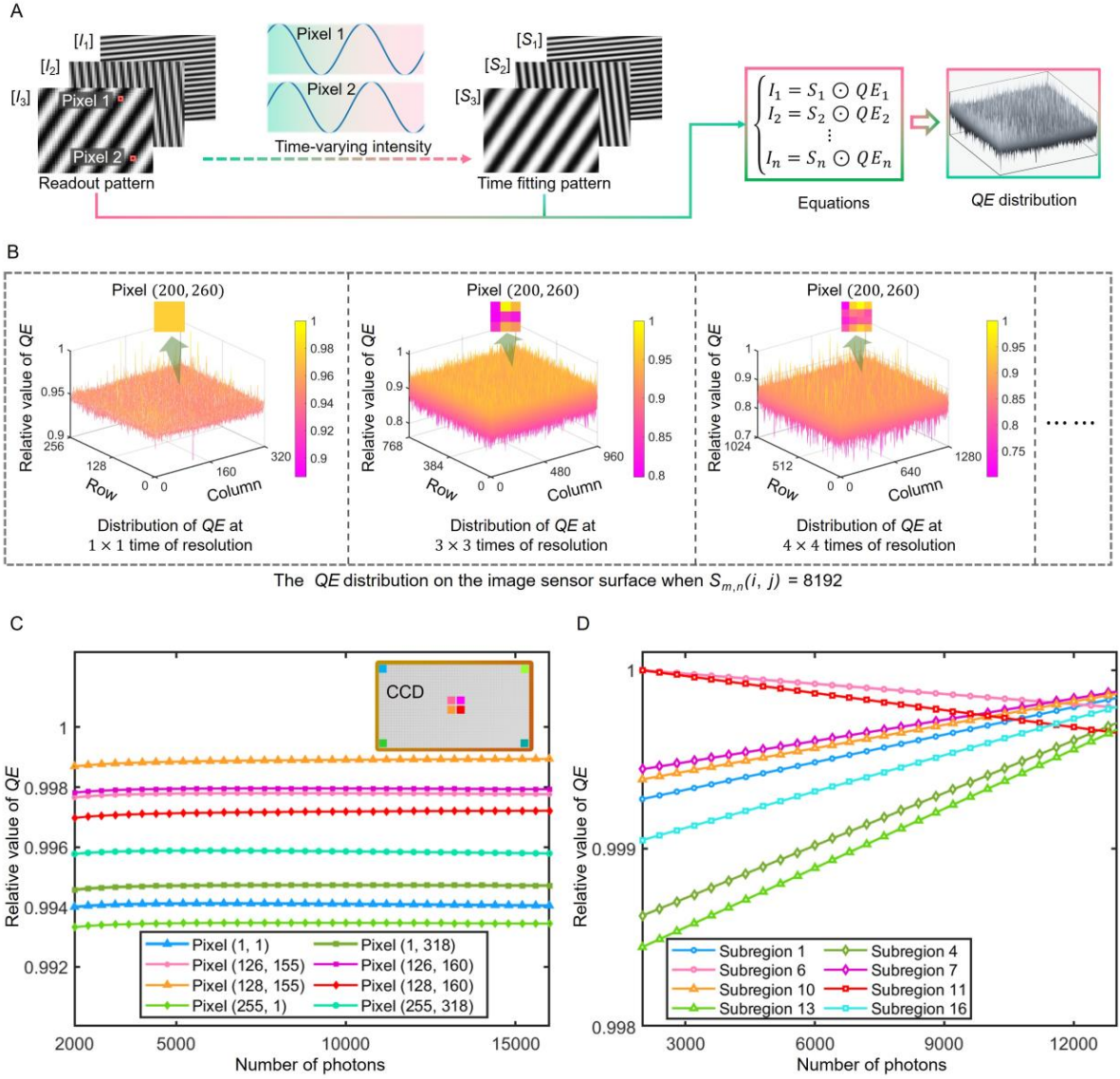

**Fig. 1 | The experimental results of *QE* measurement.** (**A**) The method to fit interference pattern and measure the *QE* distribution. (**B**) The measured *QE* distribution for the situations that each pixel is divided into 1 × 1, 3 × 3 and 4 × 4 subregions when input photo number $S_{m,n}(i, j) = 8192$ (half of the saturation intensity), in which the insets represent the intra-pixel *QE* of pixel (200, 260). (**C**) The relative *QE* values of selected pixels and subregions under different incident photon numbers, in which the inset of each figure shows their positions.

**Experimental realization and evaluation of hyper-sampling imaging**

To realize hyper-sampling imaging and evaluate its ability, we use the above CCD (pixel size = 20 μm) together with a lens system (WP-10M1220-C, WorkPower, China, $f$ = 12 mm, aperture diameter = 6 mm) to take images for a resolution plate at the distance of 0.63 m. The imaging resolution of the WP-10M1220-C lens system is about 140 lp/mm according to its data sheet. Based on the pixel size and Nyquist sampling law, the cut-off resolution of CCD is about 25 lp/mm, which means that the theoretical resolution of this whole imaging system is 25 lp/mm at most. We put the resolution plate onto a displacement platform and use the imaging system to take more than $k^2$ images $I'_{k^2}$ while it is moving. With the measured $QE_{m,n}(i, j)$ and a series of images $I'_{k^2}$ of the resolution plate, we can build at least $k^2$ equations to re-compute the input photon distribution $S'_{m,n}(i,j)$, as shown in Fig. 2A. It is obvious that the re-computed images $S'_{m,n}(i,j)$ for $k$ = 2, 3, 4 have higher resolution and better quality than the original image $I'$.

To clearly compare the difference of these images, we draw the intensity curve along the yellow line in Fig. 2A which is perpendicular to the line pairs of the resolution plate at 60 lp/mm, as shown in Fig. 2B. It can be seen that the curve of HSI (4 × 4) has the best contrast ratio than others. The ones of HSI (3 × 3) and HSI (2 × 2) are worse, but still much better than the curve of the original image. Besides that, the peak positions of HSI (4 × 4) almost overlaps with the ones of ground truth, while the peak positions of other curves deviate with ground truth a lot, which means that these images contain much less object information than HSI (4 × 4).

To be quantitative, we use modulation transfer function (MTF) to evaluate the intensity contrast at certain spatial frequency[30,31] and structural similarity index measure (SSIM) to assess the structural similarity of image to object[32,33], as shown in Fig. 2C and D. It can be seen that both the MTF and SSIM values of the original image drop dramatically to near zero after the cut-off frequency since the Nyquist sampling law can't be satisfied anymore, while the values of HSI images could stay high up to 70 lp/mm. Especially, the MTF and SSIM values of 4 × 4 HSI at the cut-off frequency are improved by 220% and 430% compared with the original image.

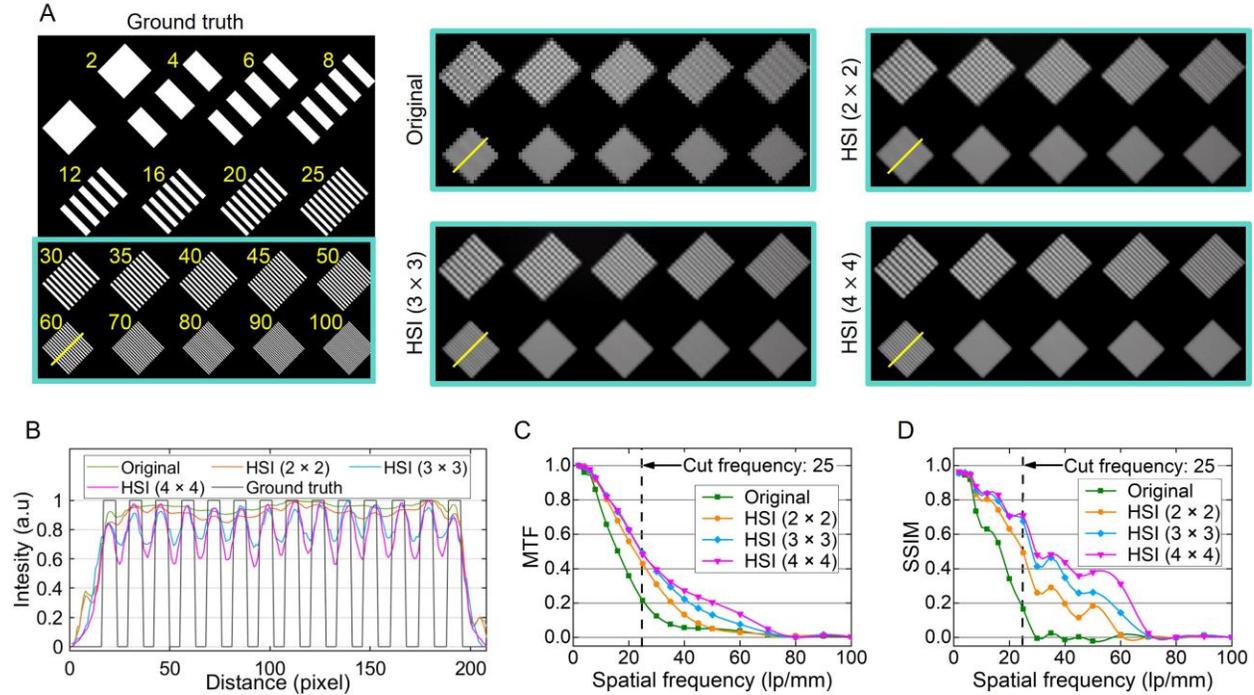

**Fig. 2 | The experimental hyper-sampling imaging results of resolution plate. (A)** The ground truth, original, HSI (2 × 2), HSI (3 × 3) and HSI (4 × 4) images of resolution plate. **(B)** The intensity curve along the yellow line in A which is perpendicular to the line pairs of 60 lp/mm on the resolution plate. **(C-D)** the MTF and SSIM values versus spatial frequency for images in A.

**Hyper-sampling imaging in practical scenes**

To show the effectiveness of HSI, we take practical images for both stationary and moving objects in laboratory and outdoor environment.

In the laboratory, we used the former Hamamatsu CCD together with a lens system (WP-5M2514-IR, WorkPower, China, f = 25.9 mm, aperture diameter = 18.5 mm) to take images for a Quick Response Code (QR code), LOGO picture, and some Chinese characters at the distance about 0.76 m. During the experiment, these objects were kept still while the camera moves along a guide rail to take more than 16 images ($k^2 = 16$). Then we implemented the HSI method to these images, and obtain the hyper-sampling image for the situation of $k = 4$. To make comparison, we also used the method of super resolution network with receptive field block based on enhanced SRGAN (RFB-ESRGAN)[16] to reconstruct high quality images. The ground truth, original, RFB-ESRGAN (16 × 16) and HSI (4 × 4) images are shown in Fig. 3A. We can visually see that the images of QR code, LOGO and Chinese characters reconstructed via HSI are much clearer than the RFB-ESRGAN and original images. To further verify the effectiveness of these methods, we used WeChat APP to recognize Chinese characters and Mean Absolute Error (MAE) to evaluate the QR code in the images. The recognized results and MAE are shown in Fig. 3B. To make quantificational evaluation, we also drew the SSIM value curves of original, RFB-ESRGAN (16 × 16), HSI (4 × 4) images at different spatial frequencies, as show in Fig. 3C. All results in Fig. 3B and C show that the HSI images have higher resolution and better quality than the ones of original and RFB-ESRGAN, which means that HSI is an imaging method that can physically retrieve objects' additional information.

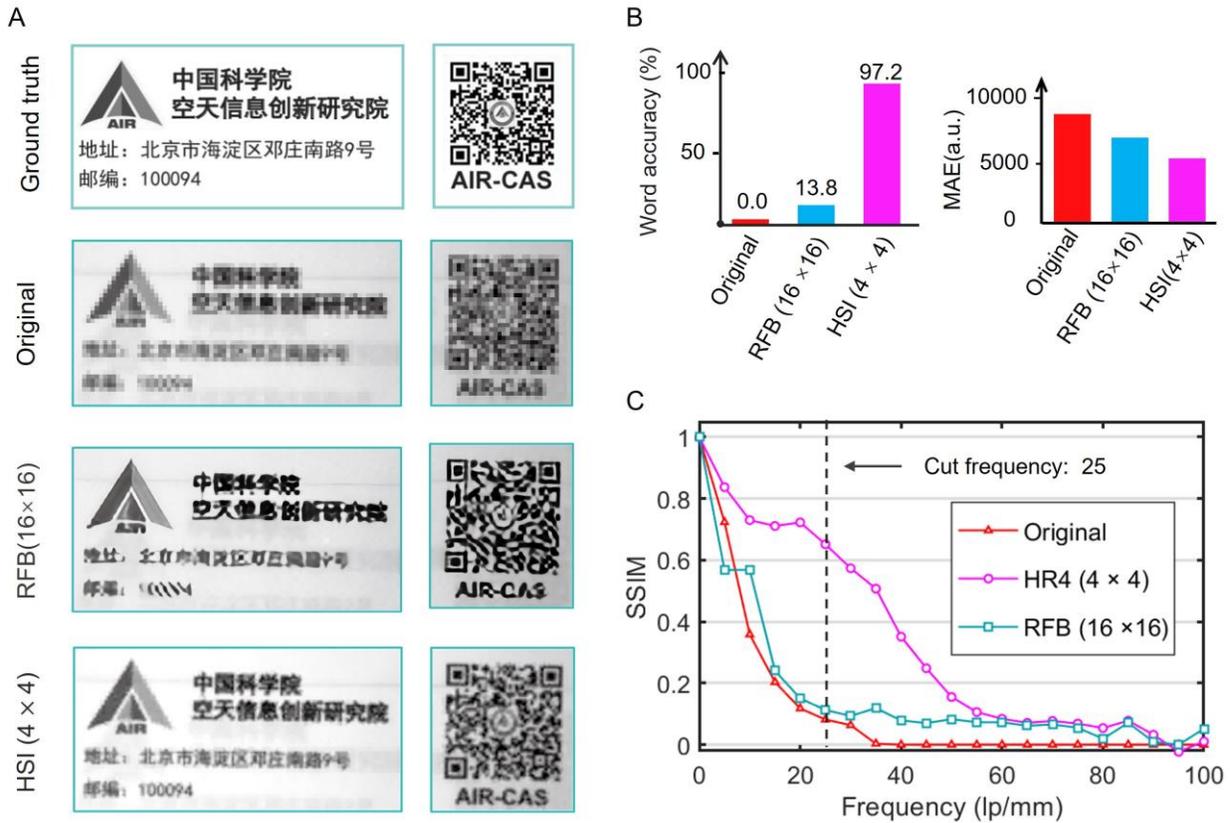

**Fig. 3 | The comparison results of different imaging methods. (A)** The imaging results of original, RFB-ESRGAN (16 × 16) and HSI (4 × 4) images to take Chinese characters and QR code. **(B)** The word accuracy and MAE values to recognize Chinese characters and QR code. **(C)** The SSIM values of original, RFB-ESRGAN (16 × 16) and HSI (4 × 4) images.

We also took images for a flying unmanned aerial vehicle (UAV) and a static building (more images of other objects are provided in the supplementary materials), which were located at 410 m and 6000 m away respectively, with the same CCD together with a lens system (600070, F/4, $f$ = 200 mm, provided by Grand Unified Optics). The lens resolution given by the manufacturer is about 150 lp/mm at the focal plane. The flying UAV (Phantom 4 RTK, DJI) is about 25 cm × 24.5 cm × 21.5 cm in size. During its flying, we used the camera to capture more than 16 images for constructing HSI images (Fig. 4A). For the static building, we moved the camera continuously during the imaging procedure to capture multiple images to achieve HSI (Fig. 4B). To quantificationally evaluate the image quality of HSI, we calculated the SSIM value at different spatial frequencies for the images under the situation of 2 × 2, 3 × 3 and 4 × 4 times sampling (the graph in Fig. 4A). The results have shown and proved that HSI method performs robustly and efficiently in practical application to physically increase the sampling resolution and imaging quality of an imaging system.

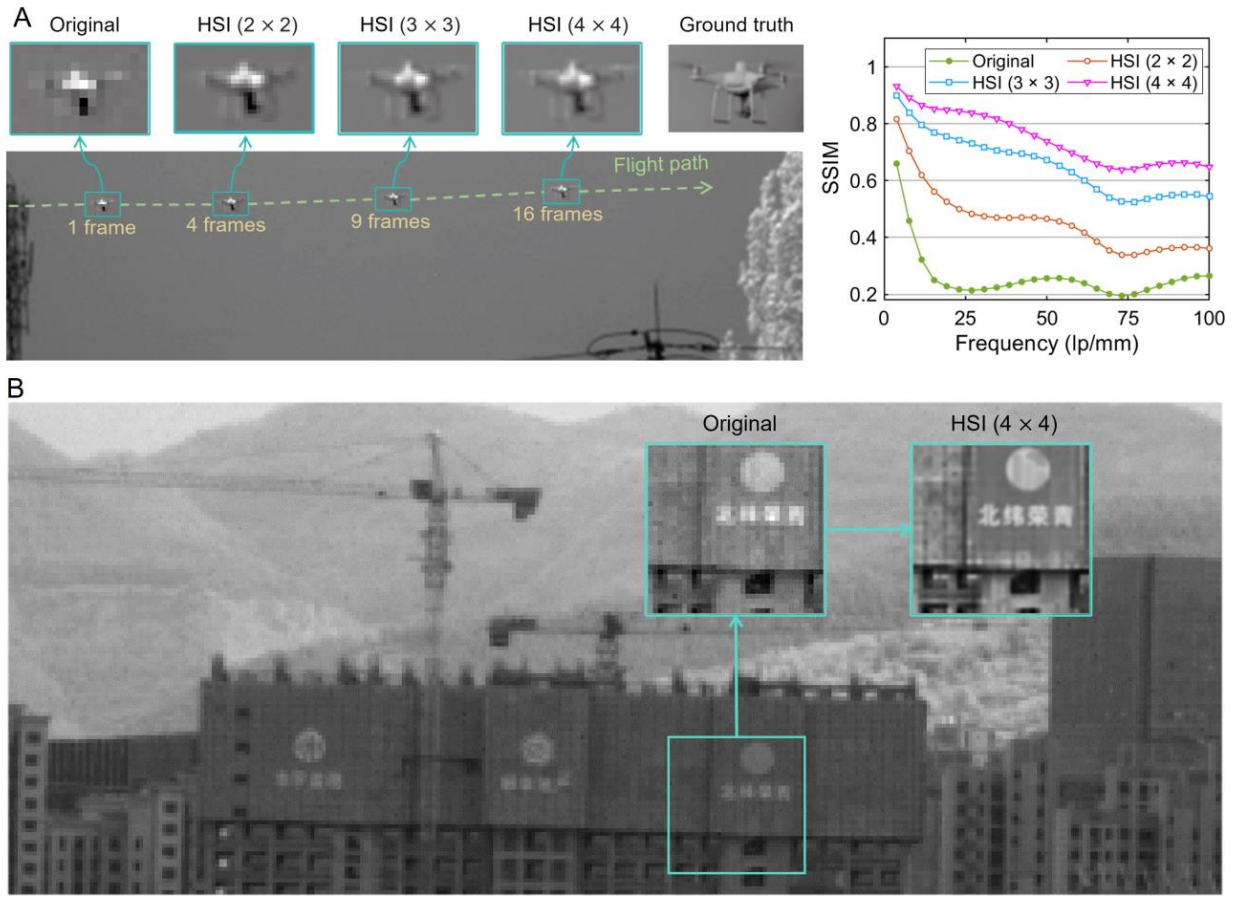

**Fig. 4 | Hyper-sampling imaging of practical objects.** (**A**) The original, HSI images of the flying UAV. As UAV flying, the HSI image becomes clearer and clearer (see movie S2). The right graph shows the SSIM curves versus spatial frequency of these images. (**B**) The original and HSI (4 × 4) images of a building.

## Discussion

In this study, we present a hyper-sampling imaging technique that significantly enhances the sampling resolution of conventional DIS through the assessment of its intra-pixel quantum efficiency and the subsequent application of image re-computation algorithms. The theoretical mechanism of HSI is illustrated at first and then verified in the laboratory by taking the images of resolution plate. Subsequently, by applying the HSI method to take images of practical objects, such as Chinese characters, QR code, flying unmanned aerial vehicle and static building, the robustness and efficiency of HSI are demonstrated and affirmed. To quantitatively and qualitatively assess the performance of HSI, we employ widely used metrics of SSIM, MTF, MAE to comprehensively evaluate the images taken by HSI and other techniques. The results reveals that images captured using the HSI technique exhibit markedly superior physical resolution and overall image quality when juxtaposed with those obtained through alternative imaging methods.

To elaborate, we believe that HSI holds the potential to markedly enhance the performance of imaging systems in three key aspects: Firstly, it offers the capacity to substantially amplify the sampling resolution of conventional sensors by many times, without incurring additional manufacturing costs or system complexity. Secondly, by reducing the required *f*-number of optical systems, HSI can lead to a significant downsizing in the physical dimensions and weight of imaging apparatuses, such as telescopes. This reduction facilitates more accessible and portable imaging solutions. Thirdly, the method is poised to notably diminish high-frequency imaging artifacts, thereby yielding a marked improvement in image clarity and fidelity.

For practical scenarios, the utilization of Hyper-Sampling Imaging (HSI) in applications can be systematically executed through a three-stage process: Initially, an optical calibration setup is employed to quantify the intra-pixel quantum efficiency distribution of the digital image sensor under time-varying laser interference patterns. This foundational stage ensures the accuracy of subsequent enhancements. The second stage involves the integration of the calibrated camera onto a mobile platform, such as satellites, aircrafts, or vehicles, to facilitate continuous imaging of stationary scenes. Alternatively, the camera can be fixed in position for the surveillance of moving targets, including aircrafts or high-speed trains. The final stage encompasses the re-computation of images utilizing the previously measured quantum efficiency distribution, culminating in the realization of HSI. Empirical evidence from both controlled laboratory environments and real-world deployments has confirmed the vast applicability of this method across a myriad of imaging domains, including terrestrial remote sensing, long-range reconnaissance, and astronomical observations[34-36].

# **Methods**

**Measurement of intra-pixel quantum efficiency**

The gray value $I_{m,n}$ of pixel $(m, n)$ is related to the total generated electron number $N_{m,n}$ of all the subregions and system gain constant $G$, in which $N_{m,n}$ is in proportion to the input photon number $S_{m,n}(i, j)$ and the performance of DIS:

$$I_{m,n} = G \cdot N_{m,n} = G \cdot \sum_{i,j=1}^{i,j=k} N_{m,n}(i,j) = \sum_{i,j=1}^{i,j=k} \left( a_{m,n}(i,j) + b_{m,n}(i,j) \cdot S_{m,n}(i,j) + c_{m,n}(i,j) \cdot S_{m,n}^2(i,j) \right) \quad (1)$$

in which, the coefficients $a_{m,n}(i, j)$, $b_{m,n}(i, j)$, $c_{m,n}(i, j)$ represent the response characteristic of subregions $(i, j)$ [37-43]

To solve $[a_{m,n}(i, j), b_{m,n}(i, j), c_{m,n}(i, j)]$, we produce slowly time-varying laser interference pattern as the input optical field (the STEP1 in Fig. 5A) to acquire sufficient amount values of $S_{m,n}(i, j)$ and corresponding $I_{m,n}$ of pixel $(m, n)$, where $I_{m,n}$ can be easily read out from the output image. Theoretically, for $k^2 \times M \times N$ subregions of a DIS with $M \times N$ pixels, we need at least $3k^2$ input optical field $S_{m,n}(i, j)$:

$$\begin{cases} I_{1,m,n} = \sum_{i,j=1}^{i,j=k} \left( a_{m,n}(i,j) + b_{m,n}(i,j) \cdot S_{1,m,n}(i,j) + c_{m,n}(i,j) \cdot S_{1,m,n}^2(i,j) \right) \\ I_{2,m,n} = \sum_{i,j=1}^{i,j=k} \left( a_{m,n}(i,j) + b_{m,n}(i,j) \cdot S_{2,m,n}(i,j) + c_{m,n}(i,j) \cdot S_{2,m,n}^2(i,j) \right) \\ \vdots \\ I_{3k^2,m,n} = \sum_{i,j=1}^{i,j=k} \left( a_{m,n}(i,j) + b_{m,n}(i,j) \cdot S_{3k^2,m,n}(i,j) + c_{m,n}(i,j) \cdot S_{3k^2,m,n}^2(i,j) \right) \end{cases} \quad (2)$$

The quantum efficiency $QE$ of each subregion can be defined as [44, 45]:

$$QE_{m,n}(i,j) = \frac{N_{m,n}(i,j)}{S_{m,n}(i,j)} = \frac{a_{m,n}(i,j)}{S_{m,n}(i,j)} + b_{m,n}(i,j) + c_{m,n}(i,j) \cdot S_{m,n}(i,j) \quad (3)$$

In the experiment, the interference pattern was formed by two laser beams $g_1(\omega)$ and $g_2(\omega + \Delta\omega)$ as the input optical field, where $\Delta\omega$ represents the frequency difference, and is set to be lower than 10 Hz in order to be well sampled by the DIS whose frame frequency is usually higher than 20 Hz (the STEP1 in Fig. 1B). By the virtue of $\Delta\omega$, the interference pattern is time-moving across sensor's plane, so we can retrieve its expression $s(x,y,t)$ in the time domain to avoid the influence of high frequency spatial display error, such as pixel size, shape, position, $QE$, etc., as shown in Fig. 1A. For subregion $(i, j)$ of pixel $(m, n)$, the received photon number $S_{\eta,m,n}(i, j)$ at time $t_\eta$ ($\eta = $ 1, 2, 3,…) can be written as:

$$S_{\eta,m,n}(i,j) = \int_{t_\eta}^{t_\eta+\tau} \int_{(m-1+\frac{i-1}{k})l_y}^{(m-1+\frac{i}{k})l_y} \int_{(n-1+\frac{j-1}{k})l_x}^{(n-1+\frac{j}{k})l_x} s(x,y,t)dxdydt \quad (4)$$

in which, $l_x$ and $l_y$ denote the length and width of each pixel; $\tau$ is digital sensor's exposure time. By subscribing incident photon number $S_{1,m,n}(i, j)$, $S_{2,m,n}(i, j)$, …, $S_{3k^2,m,n}(i,j)$ of different time into Eq. 2, the coefficients $[a_{m,n}(i, j), b_{m,n}(i, j), c_{m,n}(i, j)]$ and the intra-pixel $QE_{m,n}(i, j)$ can be solved.

**Hyper-sampling imaging**

In real imaging process, to solve the input photon distribution $S'_{m,n}(i,j)$ of the same scene for all the sub-regions, we need acquire $k^2$ gray values $I'_{k^2}$ of pixels for the same scene to build $k^2$ equations(the STEP2 in Fig. 5C):

$$\begin{cases} I_1 = \sum_{i,j=1}^{i,j=k} \left( a_1(i,j) + b_1(i,j) \cdot S'_{m,n}(i,j) + c_1(i,j) \cdot S'^2_{m,n}(i,j) \right) \\ I_2 = \sum_{i,j=1}^{i,j=k} \left( a_2(i,j) + b_2(i,j) \cdot S'_{m,n}(i,j) + c_2(i,j) \cdot S'^2_{m,n}(i,j) \right) \\ \vdots \\ I_{k^2} = \sum_{i,j=1}^{i,j=k} \left( a_{k^2}(i,j) + b_{k^2}(i,j) \cdot S'_{m,n}(i,j) + c_{k^2}(i,j) \cdot S'^2_{m,n}(i,j) \right) \end{cases} \quad (5)$$

In Eq. 5 $S'_{m,n}(1, 1)$, $S'_{m,n}(1, 2)$, …, $S'_{m,n}(k, k)$ are corresponding to the subregions of pixel($m, n$), which means that each pixel is used like $k^2$ pixels, so called hyper-sampling. In real imaging process, one way to establish Eq. 5 is moving CCD or the whole camera for the same scene during the imaging process; another way is taking images for moving objects (Fig. 5C).

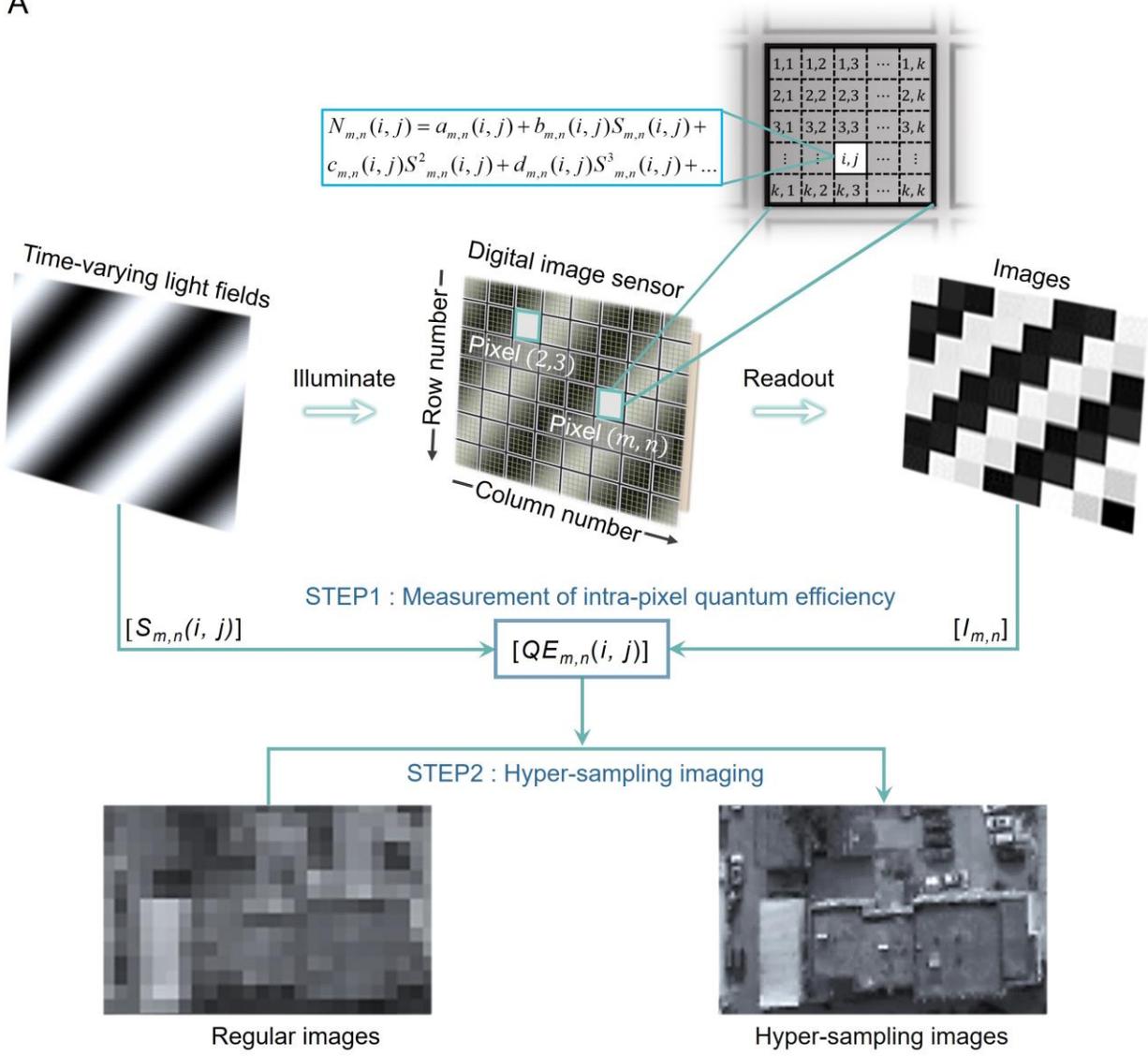

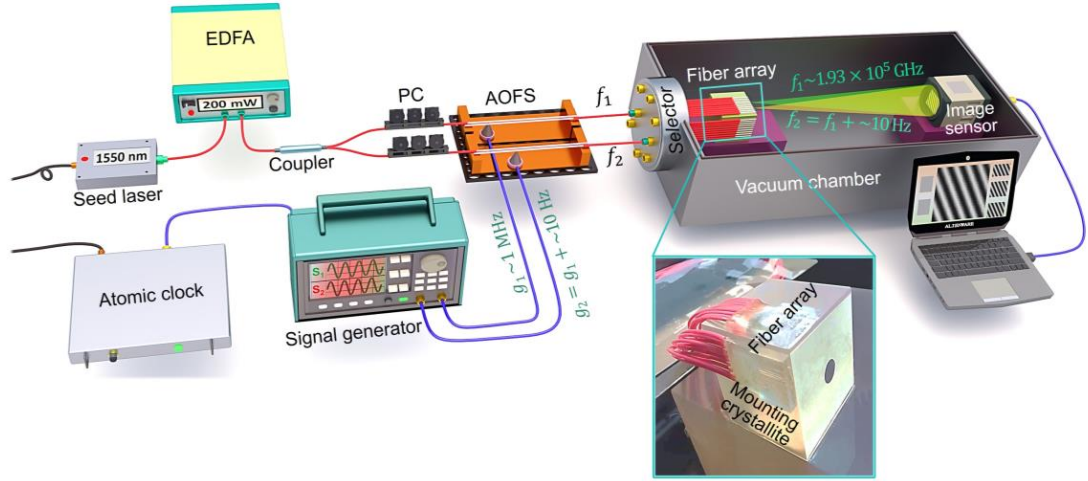

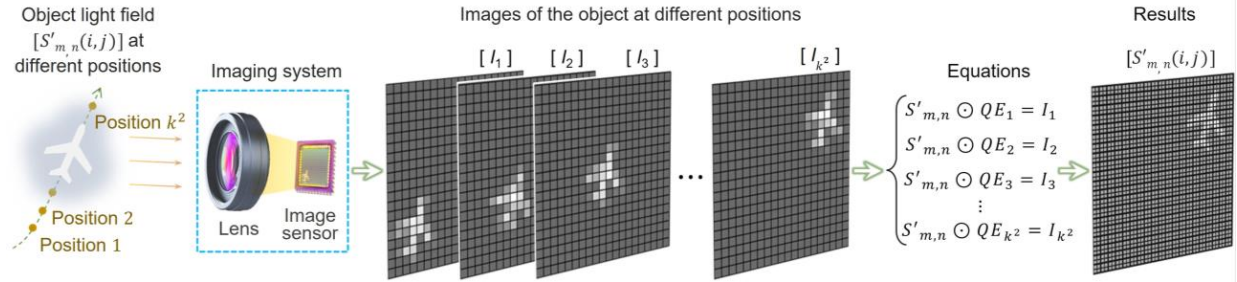

**Fig. 5 | The mechanism of intra-pix**el ***QE* measurement and hyper-sampling imaging. (A)** The two steps to realize hyper-sampling imaging. **(B)** The experiment diagram to obtain intra-pixel *QE*. EDFA: erbium doped fiber amplifier, Coupler: 50:50 coupler, AOFS: acousto-optic frequency shifter, PC: polarization controller. Detailed experiment description is provided in the supplemental materials. **(C)** The mechanism to realize hyper-sampling imaging for a moving target.

# Data and materials availability

All data needed to evaluate the conclusions in the paper are present in the paper or the supplementary materials. All the data are available from the corresponding authors upon reasonable request.

# **Acknowledgments**

We acknowledge valuable advices from Dr. Zhigang Chen, Xiang and Zhou. Ze Zhang acknowledges inspirational guidance from his previous supervisor Dr. Christodoulides.

Financial supports from the National Natural Science Foundation of China grant 61475161 (ZZ), National Natural Science Foundation of China grant 12074350 (LG), Conditional Construction

Fund of Aerospace Information Research Institute grant Y70X08A1HY (ZZ) are gratefully acknowledged.

# Author Contributions

Z.Z. designed the experiment. Z.Z., H.X, and M.S. jointly developed the theoretical model, and H.X, and M.S. completed the compilation and verification of the image processing algorithm. Z.Z., H.X, and M.S. built the optical system. Z.Z. wrote the paper with the inputs of all authors, M.G., C.S., F.G., L.G., Z.Y. J.L. and S.W. revised the paper. H.Y., H.X., M.S., J.L. and H.W carried out the specific experiments.

# Competing interests

The authors declare no competing interests.

# Additional information

**Supplementary Materials**

Materials and Methods

Figs. S1 to S5

Table S1

Movies S1 to S2